\documentclass{ws-p10x7}

\begin{document}

\title{The Weak OPE and Dimension-eight Operators}

\author{E. Golowich}

\address{Physics Department, University of Massachusetts, 
Amherst MA 01003, USA\\E-mail: golowich@physics.umass.edu}

\twocolumn[\maketitle\abstract{
We discuss recent work which identifies a potential flaw 
in standard treatments of weak decay amplitudes, including 
that of $\epsilon'/\epsilon$.  The point is that (contrary 
to conventional wisdom) dimension-eight operators 
contribute to weak amplitudes at order $G_{F} \alpha_s$ 
and without $1/M_W^2$ suppression.  The effect of dimension-eight 
operators is estimated to be at the 100$\%$ level in a sum rule 
determination of the operator 
${\cal Q}_7^{(6)}$ for $\mu = 1.5$ GeV, suggesting that presently 
available values of $\mu$ are too low to justify the neglect 
of these effects.}]

\section{Motivation}

\subsection{Calculating Kaon Weak Amplitudes}\label{subsec:OPE}

The modern approach to calculating a kaon weak nonleptonic amplitude 
${\cal M}$ involves use of the operator product expansion, 
\begin{equation}
{\cal M}  =  \sum_d~\sum_i\ {\cal C}_i^{(d)} (\mu) ~
\langle {\cal Q}_i^{(d)} \rangle_\mu \ \ , 
\label{eq:sp}
\end{equation}
in which the nonleptonic weak hamiltonian ${\cal H}_{\rm W}$ 
is expressed as a linear combination of local operators 
${\cal Q}_i^{(d)}$.  There is a sum over the dimensions 
(starting here at $d=6$) of the local operators and a sum 
over all operators of a common dimension.  In practice, the 
following hybrid methodology is employed:
\begin{enumerate}
\item The Wilson coefficients ${\cal C}_i^{(d)} (\mu)$ are 
calculated in ${\overline {MS}}$ renormalization.  
\item The operator matrix elements $\langle {\cal Q}_i^{(d)}
\rangle_\mu$ are calculated in cutoff renormalization at the 
energy scale $\mu$.  The term `cutoff' means specifically that $\mu$ 
serves as a `separation scale' which distinguishes between 
short-distance and long-distance physics.  Three different 
approaches falling into this category are quark models, 
$1/N_c$ expansion methods, and lattice-QCD 
evaluations.\footnote{A list of references is given elsewhere.\cite{dim8}}
\end{enumerate}
The reason for this hybrid approach is that it is not 
practical to carry out the (low energy) kaon matrix element 
evaluations with ${\overline {MS}}$ renormalization.  Typical 
choices for the scale $\mu$ fall in the range $0.5 \le \mu 
({\rm GeV}) \le 3$, the lower part used in quark-model and 
$1/N_c$ evaluations and the upper part in lattice simulations.  

The purpose of this talk is to describe some recent 
results:\cite{dim8}  
\begin{enumerate}
\item In a pure cutoff scheme, dimension-eight operators occur 
in the weak hamiltonian at order $G_{F} \alpha_s / \mu^2$, $\mu$ 
being the separation scale. This can be explicitly demonstrated 
(see Sect.~2) in a 
calculation involving a LR weak hamiltonian. 
\item In dimensional regularization (DR), the $d=8$ operators do 
{\it not} appear explicitly in the hamiltonian at order $G_{F} 
\alpha_s$. However, the use of a cutoff scheme 
for the calculation of the matrix elements of
dimension-six operators requires a careful matching onto DR 
for which dimension-eight operators {\it do} play an important role.  
\end{enumerate}

These findings mean that hybrid evaluations, in the sense described 
above, of kaon matrix elements at low $\mu$ will contain (unwanted) 
contributions from dimension-eight operators.  At the very least, 
this will introduce an uncertainty of unknown magnitude 
into the evaluation.   

\section{Cutoff Renormalization}

\subsection{$\epsilon'/\epsilon$ in the Chiral 
Limit}\label{subsec:epsilon}
The determination of $\epsilon'/\epsilon$ can be shown 
to depend upon the matrix elements 
$\langle(\pi\pi)_0|{\cal Q}_6^{(6)}|K\rangle$ and 
$\langle(\pi\pi)_2|{\cal Q}_8^{(6)}|K\rangle$.\cite{buras}  In the 
chiral limit of vanishing light-quark mass, the latter 
matrix element (as well as that of operator ${\cal Q}_7^{(6)}$)
can be inferred from certain vacuum expectation values, 
$\langle 0|{\cal O}_{1,8}^{(6)}|0\rangle \equiv 
\langle {\cal O}_{1,8}^{(6)}\rangle $, where 
${\cal O}_{1,8}^{(6)}$ are dimension-six four-quark 
operators.\cite{dg}  The use of soft-meson 
techniques to relate physical amplitudes to those in the world of 
zero light-quark mass is a well-known procedure of chiral dynamics.  

\subsection{Sum Rules for $\langle 
{\cal O}_{1,8}^{(6)}\rangle$}\label{subsec:SR}
Numerical values for $\langle {\cal O}_{1,8}^{(6)}\rangle$ 
in cutoff renormalization can be obtained from the 
following sum rules,\cite{dg}
\begin{eqnarray}
& & {16 \pi^2 \over 3} \langle {\cal O}_1^{(6)}\rangle_\mu^{\rm (c.o.)} 
= \int_0^\infty ds~s^2 \ln {s + \mu^2 \over s} 
\Delta \rho \nonumber \\
& & 2 \pi \langle \alpha_s {\cal O}_8^{(6)}\rangle_\mu^{\rm (c.o.)} 
= \int_0^\infty ds~s^2 {\mu^2 \over s + \mu^2 } 
~\Delta \rho  \ , \nonumber \\
\label{sr} 
\end{eqnarray} 
where $\Delta \rho (s)$ is the difference of vector and axialvector 
spectral functions, and $\Delta \Pi (Q^2)$ is the 
corresponding difference of isospin polarization functions 
(${\cal I}m ~\Delta \Pi = \pi \Delta \rho$).  

\subsection{Physics of a LR Operator}\label{subsec:LR}

One can probe the influence of $d=8$ operators by considering 
the K-to-$\pi$ matrix element ${\cal M}(p)$, 
\begin{eqnarray}
{\cal M} (p) &=& \langle \pi^- (p) | {\cal H}_{\rm LR} | K^-
(p) \rangle \ \ , 
\label{amp}
\end{eqnarray}
where ${\cal H}_{\rm LR}$ is a LR hamiltonian obtained by 
flipping the chirality of one of the quark pairs in the usual 
LL hamiltonian ${\cal H}_{\rm W}$.  The reason for defining such 
a LR operator is that, in leading chiral order, its K-to-$\pi$ 
matrix element is nonzero and yields information on 
$\langle {\cal O}_{1}^{(6)}\rangle$ and 
$\langle {\cal O}_{8}^{(6)}\rangle$. 

To demonstrate this, we proceed 
to the chiral limit to find 
\begin{eqnarray}
& & {\cal M} \equiv {\cal M}(0) = \lim_{\rm p = 0} {\cal M} (p) 
\nonumber \\
& & = {3 G_F M_W^2 \over 32 \sqrt{2}\pi^2 
F_\pi^2} \int_0^{\infty} dQ^2 \ {Q^4 \over Q^2 + M_W^2} 
~\Delta \Pi \ . \nonumber \\
\label{chir}
\end{eqnarray}
This result is {\it exact} --- it is not a consequence of 
any model.  Information about 
$\langle {\cal O}_{1}^{(6)}\rangle$ and 
$\langle {\cal O}_{8}^{(6)}\rangle$ is obtained by 
performing an operator product expansion on 
$\Delta \Pi(Q^2)$.  Working to first order in $\alpha_s$ we have 
\begin{eqnarray}
& & {\cal M} = {G_F \over 2 \sqrt{2}F_\pi^2}
\bigg[ \langle {\cal O}_1^{(6)} \rangle_\mu^{\rm (c.o.)} 
\nonumber \\
& & + { 3 \over 8 \pi} \ln {M_W^2 \over \mu^2}
\langle \alpha_s {\cal O}_8^{(6)} \rangle_{\mu} 
+ {3\over 16 \pi^2}{{\cal E}^{(8)}_\mu\over \mu^2} + 
 \ldots \bigg] \nonumber \\
\label{e8}
\end{eqnarray}
The three additive terms in Eq.~(\ref{e8}) are proportional 
respectively to the quantities $\langle {\cal O}_{1}^{(6)}\rangle$,  
$\langle {\cal O}_{8}^{(6)}\rangle$ and 
${\cal E}^{(8)}$.  The last of these (${\cal E}^{(8)}$) contains the 
effect of the $d=8$ contributions.\footnote{Although the $d=8$ 
LL operators arising from ${\cal Q}_2^{(6)}$ have been 
determined\cite{dim8}, to our knowledge the individual $d=8$ LR 
operators comprising ${\cal E}^{(8)}$ have not.}   For dimensional 
reasons, ${\cal E}^{(8)}$ must be accompanied by an inverse squared 
energy.  This turns out to be the factor $\mu^{-2}$.  

In Table 1 we display the numerical values 
(in units of $10^{-7}~{\rm GeV}^2$) of the three 
terms of Eq.~(\ref{e8}) for various choices of $\mu$.  
Observe for the lowest values that the dimension-eight 
term dominates the contribution from 
$\langle {\cal O}_{1}^{(6)}\rangle$.  
Only when one proceeds to a sufficiently large value like 
$\mu = 4$~GeV is the $d=8$ influence suppressed.

\begin{table}
\caption{Eq.~(\ref{e8}) in units of $10^{-7}~{\rm GeV}^2$.}\label{tab:smtab}
\begin{tabular}{|c||c|c|c|} 
 
\hline 
 
\raisebox{0pt}[12pt][6pt]{$\mu$~(GeV)} & 
 
\raisebox{0pt}[12pt][6pt]{Term 1} & 
 
\raisebox{0pt}[12pt][6pt]{Term 2} &
 
\raisebox{0pt}[12pt][6pt]{Term 3} \\
 
\hline
 
\raisebox{0pt}[12pt][6pt]{$1.0$} & 
 
\raisebox{0pt}[12pt][6pt]{$-0.12$} & 
 
\raisebox{0pt}[12pt][6pt]{$-3.84$} & 
 
\raisebox{0pt}[12pt][6pt]{$0.64$} \\

\hline

\raisebox{0pt}[12pt][6pt]{$1.5$} & 
 
\raisebox{0pt}[12pt][6pt]{$-0.28$} & 
 
\raisebox{0pt}[12pt][6pt]{$-3.49$} & 
 
\raisebox{0pt}[12pt][6pt]{$0.30$} \\

\hline

\raisebox{0pt}[12pt][6pt]{$2.0$} & 
 
\raisebox{0pt}[12pt][6pt]{$-0.44$} & 
 
\raisebox{0pt}[12pt][6pt]{$-3.24$} & 
 
\raisebox{0pt}[12pt][6pt]{$0.17$} \\

\hline

\raisebox{0pt}[12pt][6pt]{$4.0$} & 
 
\raisebox{0pt}[12pt][6pt]{$-0.89$} & 
 
\raisebox{0pt}[12pt][6pt]{$-2.63$} & 
 
\raisebox{0pt}[12pt][6pt]{$0.04$} \\\hline
\end{tabular}
\end{table}

\section{Dimensional Regularization}
Suppose one wishes to express the entire analysis in terms of 
${\overline {MS}}$ quantities.  To do so requires converting 
matrix elements in cutoff renormalization to those in 
${\overline {MS}}$ renormalization.   Recall, in 
dimensional regularization one calculates in $d$ dimensions and 
for dimensional consistency introduces a scale $\mu_{\rm d.r.}$. 

The dimensionally regularized matrix element for 
$\langle {\cal O}_1^{(6)}\rangle$ is found from 
the $d$-dimensional integral,\cite{dg} 
\begin{eqnarray}
& & \phantom{xxxx} \langle {\cal O}_1^{(6)}\rangle_\mu^{\rm (d.r.)} = 
\langle {\cal O}_1^{(6)}\rangle_\mu^{\rm (c.o.)} 
\nonumber \\
& & + {d - 1 \over (4 \pi)^{d/2}} 
{\mu_{\rm d.r.}^{4-d} \over \Gamma \left(d/2 \right)} 
\int_{\mu^2}^{\infty} dQ^2 \ Q^d ~\Delta \Pi (Q^2)\ . \phantom{xxxx} 
\label{dr} 
\end{eqnarray} 
The term in Eq.~(\ref{dr}) containing the integral is proof 
that the dimensionally regularized matrix element  
$\langle {\cal O}_1^{(6)}\rangle_\mu^{\rm (d.r.)}$ will contain 
{\it short-distance} contributions.  As written, this term 
becomes divergent for four dimensions and also is scheme-dependent.  
In the ${{\overline {\rm MS}}}$ approach, 
the divergent factor $2/\epsilon - \gamma + \ln (4 \pi)$ is 
removed.  The NDR scheme involves a certain procedure for 
treating chirality in $d$-dimensions.  The final result 
is a relation (given here to ${\cal O}(\alpha_s)$) between the 
cutoff and ${\overline{\rm MS}}$-NDR matrix elements,  
\begin{eqnarray}
& & 
\langle {\cal O}_1^{(6)}\rangle_\mu^{\rm ({\overline {MS}}-NDR)} = 
\langle {\cal O}_1^{(6)}\rangle_\mu^{\rm (c.o.)} 
\nonumber \\
& & + {3 \over 8 \pi } 
\left[ \ln {\mu_{\rm d.r.}^2 \over \mu^2} 
- {1 \over 6} \right] \langle \alpha_s {\cal O}_8^{(6)} \rangle_\mu 
\nonumber \\
& & + {3 \over 16 \pi^2} \cdot {{\cal E}^{(8)}_\mu \over \mu^2} + 
\dots 
\label{codr}
\end{eqnarray} 
The effect of the $d=8$ contribution to the weak OPE 
now appears in the $d=6$ ${\overline{\rm MS}}$-NDR operator  
matrix element.  Note also that the parameter 
$\mu_{\rm d.r.}$ is distinct from the separation scale $\mu$. 

\section{Evaluation of $B_{7,8}^{(3/2)}$}
To suppress the effect of dimension-eight operators on 
the determinations of Eq.~(\ref{sr}), one should 
evaluate the two sum rules for $\langle {\cal O}_{1,8} 
\rangle_\mu^{\rm (c.o.)}$ at a large value of $\mu$ ({\it e.g.} 
$\mu \ge 4$~GeV) and then use renormalization group 
equations to run the matrix elements down to lower values of $\mu$ 
({\it e.g.} $\mu = 2$~GeV).\cite{dgh}  Alternative approaches 
might involve the finite energy sum rule framework\cite{gm} or 
QCD-lattice simulations at sufficiently large $\mu$.

\section{Concluding Remarks}
This talk has dealt with an important aspect of 
calculating kaon weak matrix elements, the role of 
dimension-eight operators.  In this regard, 
Eq.~(\ref{codr}) is of special interest. It reveals 
that the relation between ${\overline{\rm MS}}$-NDR and 
cutoff matrix elements will involve not only 
mixing between operators of a given dimension 
but also mixing between operators of differing dimensions.  
The net result of our work is that 
existing work on $\epsilon' /\epsilon$ will be affected, 
especially for methods which take $\mu \le 2$~GeV.  

\section*{Acknowledgments}
This work was supported in part by the National
Science Foundation.

\end{document}